\documentstyle[12pt]{article}
\newcommand{\tr}[1]{\,{\rm tr}\,#1\,}

\begin{document}
\title{
\begin{flushright}
{\small SMI-4-93 }
\end{flushright}
\vspace{0.5cm}
Uniqueness of $U_q(N)$ as a quantum gauge group\\ and\\
representations of its differential algebra}
\author{ I.Ya.Aref'eva
\\
 and\\
 G.E.Arutyunov \thanks{E-mail:$~~$ arefeva@qft.mian.su, arut@qft.mian.su}
\thanks{Supported in part by RFFR under grant N93-011-147 }
\\ Steklov Mathematical Institute, Russian Academy of Sciences,\\
Vavilov st.42, GSP-1,117966, Moscow, Russia }
\maketitle
\begin{abstract}
To construct a quantum group gauge theory one needs
an algebra which is  invariant under gauge transformations.
The existence of this invariant algebra is closely related with
the existence of a differential algebra $\delta _{{\cal H}} G_{q}$
compatible with the Hopf algebra
structure. It is shown that $\delta _{{\cal H}} G_{q}$
exists only for the quantum group $U_{q}(N)$ and
that the quantum group $SU_q(N)$
as a  quantum gauge group is not allowed.

The representations of the algebra $\delta _{{\cal H}} G_{q}$ are constructed.
The operators corresponding to the
differentials are realized
via derivations on the space of all irreducible
${*}$-representations of $U_q(2)$.
With the help of  this construction   infinitesimal gauge transformations
in two-dimensional classical space-time are described.
\end{abstract}

\newpage
\section   {Introduction}
Recently a construction of quantum group gauge theory (QGGT), i.e. the
gauge theory with a quantum group playing the role of the gauge group, has
been initiated \cite {AV}-\cite {IP1}.

In spite of the impressive successes
of applying gauge theory to the description of all known physical
interactions the natural question about the possibilities of extending the
strict frames of gauge theories arises. One can think  that  an enlargement of
the rigid framework of gauge theory would help to solve  fundamental
theoretical problems of spontaneous symmetry breaking and quark confinement.
The theory of quantum groups looks rather attractive as the mathematical
foundation of a new theory since the general requirements of symmetries of
a physical
system can be formulated on the language of quantum groups
\cite{Bie,Ar}.

At the last two years attempts where made to understand the algebraic structure
of QGGT \cite{AV}-\cite{IP1}. The main efforts have been to keep the
classical form of  gauge transformations for the gauge potential $A$:
\begin{equation}
A\rightarrow
A'=TAT^{-1}+dTT^{-1}.  \label{0}
\end{equation}
Roughly speaking, the
problem is in the  following. Assume that $T$ is an element of a quantum group.
What differential calculus should we consider and from what algebra ${\cal A}$
should be taken a gauge potential $A$ to guarantee that
$A'$ also belongs to the algebra ${\cal A}$?
One of suitable resolutions of this problem has been recently found by Isaev
and Popovicz \cite{IP1}. In their scheme $T$ and
$dT$ are realized as generators of the differential extension
$\delta _{{\cal H }}G_{q}$  of a quantum group
$G_{q}$ compatible with the Hopf algebra structure.

It is natural to try to find
$\delta _{{\cal H}} $-extension for $SU_q(N)$ that would lead to the
algebraic formulation of the $SU_q(N)$ gauge theory. It is known that the
$\delta _{{\cal H}} $ extension  of $GL_q(N)$ does exist \cite{WZ,IP1,IPy}.
In this paper we deal with
the analogous construction for the quantum group $U_q(N)$.
The quantum group $U_q(N)$ is one
of the real forms of $GL_q(N)$ and may be obtained from $GL_q(N)$ by
introducing ${*}$-involution operation. The $\delta _{{\cal H}} $ extension of
$GL_q(N)$ also admits ${*}$-involution and we get $\delta _{{\cal H}}(U_q(N))$.
To  obtain the $\delta _{{\cal H}}$-extension of $SU_q(N)$ one needs to fix the
quantum
determinant equal to unity.  However, at  this point
one obstacle arises. Namely, {\it the quantum determinant is not
a central element} of $\delta _{{\cal H}}(U_q(N))$ and therefore
cannot be fixed equal to unity.

Summing up, it is possible to present an algebraic construction of the
quantum group gauge potential for $U_q(N)$ but the quantum group $SU_q(N)$
as a gauge group is not allowed.
Thus, if one believes
that the ordinary gauge theory is obtained as the classical limit,
$q\rightarrow 1$, of some QGGT then one can speculate that QGGT predicts
the group $U(2)$. From this point of view the fact that electroweak group is
$U(2)=SU(2)\otimes U(1)$ looks rather promising.

Fields defined on the classical space-time and taking
value on a quantum group
or quantum algebra should be the natural object of the QGGT
\cite{AV}.  However, just at this point there is the
problem for a straightforward application of the standard approach to
quantum groups. The ordinary local gauge theory is based on the existence of
a sufficiently wide class of differentiable maps from
space-time into a group. This class can be easily constructed since
a Lie group is a smooth manifold and so it is possible to regard
its coordinates as functions on space-time. In the standard  quantum group
approach the space of c-number parameters numerating points of a quantum
group is not available.
Usually the theory of quantum groups is formulated in terms of the
function algebra $Fun(G_q)$ on a quantum group $G_q$.  Adopting this view
and trying to describe QGGT one can expect that ordinary
gauge theory may be formulated in terms of the function algebra $Fun(G_q)$,
i.e. on the dual language. However, it is not suitable for field theory
applications. So it is clear that one cannot build QGGT in the framework
of the standard quantum group approach. We need to extend the usual
content of quantum groups by introducing in the theory new objects. In
other words, to consider a map of the classical space-time $R^4$ into a
quantum group we need a more liberated treatment of a quantum  group or quantum
plane than  the ordinary theory offers. Such an approach was suggested in
\cite {AV}.
For consideration of this problem see also  \cite {AVII,Luk,Koz,Rim,Ubr}.
An example showing the necessity of introduction new  objects
is given by the analogue of the exponential map
for quantum groups. It turns out that in addition to a
quantum group one should introduce a set of
generators taking value in so-called "quantum superplane" \cite{AV}.

In this paper we  present
an explicit realization of a differentiable local map from the classical
space-time $R^{2}$ into the quantum group $U_q(2)$ supplied with
$*$-involution and compatible with the bicovariant differential
calculus on $U_q(2)$. We will see that for this purpose
it is suitable to consider a quantum group as the set of all its
irreducible unitary representations and think of parameters numerating
these representations as "coordinates" on a quantum group. Note that this
consideration is in the line of the approach  \cite {AV}
(see also \cite{Luk}) and has an implicit
support in a definition  of integral on a quantum group
proposed in \cite{Wor}\footnote{This approach to integral on quantum group
is used to define a lattice QGGT  \cite {AAL}}.
By the compatibility of a map
$R^2\rightarrow U_q(2)$ with the bicovariant differential calculus
on $U_q(2)$ we mean that exterior derivative $d$
acting on elements of the quantum group
can be decomposed over a basis $\{ \varepsilon ^i \}$ of ordinary
differential forms on $R^{2}$:
$d=\varepsilon ^i\otimes \partial _i.$

To get representations for the derivatives $\partial _i$ we start from
a construction of representations for $(T,L)$-pair, where
$L$ is a quantum gauge field with zero curvature $L=dTT^{-1}=\left(
\begin{array}{cc}\omega _{0} & \omega _{+} \\
\omega _{-} & \omega _{1}\end{array}\right)~$. The simple consideration
shows that it is impossible to realize the operators
$\omega _{0}, \omega _{+}, \omega _{-}, \omega _{1}$ in the space of
an irreducible representation of the algebra $Fun(U_{q}(2))$. This means that
we have to extend the space of representation. We deal with a direct
integral of Hilbert spaces over parameters labelled irreducible representations
of $U_{q}(2)$. We find a simple formulae for the operators
$\omega _{0}, \omega _{+}, \omega _{-}, \omega _{1}$  and then get the
representations  for $dT$. It turns out that
there are two types of representations of $dT$ corresponding to
differentiations over two parameters
specified irreducible unitary representations of $U_q(2)$. For these
two different
representations we use notations $\partial _1T$ and $\partial _2T$.
Thus we realize the derivatives $\partial _i$ as differentiations over
parameters
of representations of a quantum group itself, i.e. as differentiations over
"coordinates" on a quantum group.

Now having at hand an explicit form of derivatives of  $U_q(2)$ elements
one can locally construct a differentiable map $R^2\rightarrow U_q(2)$ as
following:
\begin{equation}
T(x)=\left(\begin{array}{ll}a(x)&b(x)\\c(x)&d(x) \end{array}\right)=
\left(\begin{array}{ll}a+x^i\partial
_i a & b+x^i\partial _ib \\ c+x^i\partial _ic &
d+x^i\partial _id  \end{array}\right), ~i=1,2.
                    \label{qq}
\end{equation}
Differentiation of operators $a(x),\ldots ,d(x)$ with respect to $x^1,x^2$
gives the derivatives compatible with the bicovariant differential
calculus.
In this construction we have to limit ourselves by considering the
two-dimensional classical space-time
since the space of parameters of infinite dimensional unitary representations
of $U_{q}(2)$ is two-dimensional.

Therefore we have for the bicovariant
differential calculus on $U_q(2)$ a usual "classical" picture: if the
quantum group is a set of its irreducible unitary representations then
the quantum group derivatives are indeed derivatives with respect to the
coordinates on $U_q(2)$.

The paper is organized as follows. In section 2 we describe an algebraic
approach to constructing QGGT. In section 3 we introduce $*$-involution
for the $\delta _{{\cal H}}$-extension of the algebra $Fun(GL_q(N))$ and
prove a no-go theorem that only $\delta _{{\cal H}}(U_q(N))$ exists and
the $\delta _{{\cal H}}$-extension of
$SU_q(N)$ is not allowed. In sections 4 and 5 two
inequivalent $*$-representations of the $\delta _{{\cal H}}$-extension of
$U_q(2)$
in the Hilbert space are constructed.  We use them in section 6 to write
out an explicit form of a two-dimensional map into the quantum group
$U_q(2)$.

\section{Algebraic Scheme of QGGT}
Let $T$ belongs to a quantum matrix group $G$,
i.e. $T$ is the subject of the relations:
\begin{equation} R_{12}T_{1}T_{2}=T_{2}T_{1}R_{12}.
\label{cl}
\end{equation}
Here $R_{12}$ is a quantum R-matrix and $T_1=T\otimes I,T_2=I\otimes T$
(see \cite{Fad,M} for details).
Recall that due to the existence of the
Hopf algebra structure for quantum groups the product $gT$ of two elements
being the subjects of equation (\ref{1}) satisfies
\begin{equation}
R_{12}(gT)_{1}(gT)_{2}=(gT)_{2}(gT)_{1}R_{12},
\label{d1}
\end{equation}
if the entries of the matrices $g$ and $T$ mutually commute.

To construct QGGT
one can start with the consideration of gauge fields
having zero curvature \cite{AVI}:
\begin{equation} L=dTT^{-1}.  \label{1} \end{equation}

In order to give a meaning to
(\ref{1}) we need to specify the differential $dT$ on a quantum group,
i.e. to determine the differential calculus.
Differential calculi on
quantum groups were developed in \cite{Fad,Wor,WZ,FAS,IPy}.
The operator of exterior derivative is supposed to have the usual
properties:
\begin{equation}
d^2=0,~~d(AB)=(dA)B+A(dB).    \label{rr}
\end{equation}

If $T$ and $dT$ are understood as matrices with non-commutative entries
then $L$ is also a matrix with non-commutative entries. It is interesting
to know if there are permutation relations between the entries of $L$ that
can be written in terms of R-matrix. For the case of $GL_q(N)$ the answer is
yes if one considers
a special differential calculus.
It can be formulated by introducing the set of
generators $dT_{ij}$ that satisfy the relations \cite{IP1,WZ,AVI,FAS}:
\begin{equation}
R_{12}(dT)_{1}T_{2}=T_{2}(dT)_{1}R_{21}^{-1},                \label{5}
\end{equation}
\begin{equation}
R_{12}(dT)_{1}(dT)_{2}=-(dT)_{2}(dT)_{1}R_{21}^{-1}.
\label{6}
\end{equation}
The relations (\ref{cl}),(\ref{5}) and the definition (\ref{1}) yield
the quadratic algebra for $L$:
\begin{equation}
R_{12}L_1R_{21}L_2+L_2R_{12}L_1R_{12}^{-1}=0.
                                                                \label{7}
\end{equation}
The differential calculus \cite{WZ,IP1} is compatible
with the Hopf
algebra structure in the sense that the following equation is satisfied
\begin{equation}
R_{12}d(gT)_{1}(gT)_{2}=d(gT)_{2}(gT)_{1}R_{21}^{-1}
\label{tq}
\end{equation}
(for more precise definition see section 3).
For ordinary groups
the field $L$ is transformed under gauge transformations \begin{equation}
T\rightarrow gT \label{2} \end{equation} as follows:
\begin{equation} L\rightarrow L'=gLg^{-1}+dgg^{-1}. \label{3}
\end{equation}

It is remarkable that equations (\ref{d1}) and (\ref{tq}) are enough to
guarantee the invariance of the algebra (\ref{7}) under transformations
(\ref{3}).  Thus to construct QGGT we have to extend
our quantum group to the $\delta _{{\cal H}}$ Hopf algebra.

The next nontrivial step is to postulate for the gauge potential $A$ of
the general form  the same quadratic algebra as for $L$:
\begin{equation}
R_{12}A_1R_{21}A_2+A_2R_{12}A_1R_{12}^{-1}=0.
                                                                \label{8}
\end{equation}
The relations (\ref{d1}),(\ref{tq}) are again enough for invariance of
(\ref{8}) under gauge transformations (\ref{0}). Note that
as in the case of zero curvature potential we assume that the entries of
the matrices $A$ and $g$ are mutually commutative. In what follows we will
regard the quadratic algebra (\ref{8}) as the algebra $\cal{A}$ of
quantum group gauge potentials. The corresponding curvature has the form
$$F=dA-A^2$$ and it is transformed under gauge transformations as
\begin{equation} F\rightarrow
gFg^{-1}. \label{9}
\end{equation}

To conclude the brief presentation of the formal algebraic QGGT
construction one has to mention that the action should be taken in the
form $tr_qF^2$, since
the $q$-trace \cite{Fad,IP} is invariant  under  (\ref{9}).
Note that $F$ also belongs to a quadratic algebra \cite{IP1}
defined by the reflection equations \cite{Kul}.

\section{{*}-involution for the $\delta _{{\cal H}}$ extension of
$Fun(U_{q}(2))$}
The $\delta _{{\cal H}}$ extension $\delta _{{\cal H}}(GL_q(N))$ of
the Hopf algebra $Fun(GL_{q}(N))$ is the Hopf algebra \cite{IP} itself with the
comultiplication $\Delta $, the counity $\epsilon $ and the antipod ${\cal S}$
which are defined by
\begin{displaymath}
\Delta (T)=T\otimes T~,~~\epsilon (T)=1~,~~{\cal S}(T)=T^{-1}~,
\end{displaymath}
\begin{equation}
\Delta (dT)=dT\otimes T+T\otimes dT~,
{}~~\epsilon (dT)=0~,~~{\cal S}(dT)=-T^{-1}dTT^{-1}~.
                                                         \label {sah}
\end{equation}

Now we are going to show that $\delta _{{\cal H}}(GL_q(N))$ admits
${*}$-involution.

Recall the definition of ${*}$-involution of a Hopf algebra.
An involution ${*}$
of a Hopf algebra ${\cal A}$ is a map ${\cal A} \rightarrow {\cal A}$
which is the algebra antiautomorphism and the coalgebra automorphism
obeying two conditions:
\begin{enumerate}
\item $(a^{*})^{*}=a$
\item ${\cal S}({\cal S}(a^{*})^{*})=a$ for any $a\in {\cal A}~.$
\end{enumerate}

Let $q$ be real.
Supposing the existence of the ${*}$-involution for the Hopf algebra ${\cal A}$
and applying ${*}$ to equation (\ref{5}) one finds:
\begin{equation}
R_{12}(T^{*})_{2}(dT^{*})_{1}=(dT^{*})_{1}(T^{*})_{2}R_{21}^{-1}
                                                         \label {ei}
\end{equation}
Using the Hopf
algebra structure of $\delta _{{\cal H}}(GL_q(N))$ (\ref {sah}) and defining
relations (\ref{cl})-(\ref{5}) one can deduce that ${\cal S}(dT)$ obeys
equation:
\begin{equation}
R_{12}({\cal S}(T)^{t})_{2}({\cal S}(dT)^{t})_{1}=
({\cal S}(dT)^{t})_{1}({\cal S}(T)^{t})_{2}R_{21}^{-1}~,
                                                         \label {ea}
\end{equation}
where $t$ means the matrix transposition. Comparing (\ref{ei}) and (\ref{ea})
we see that it is possible to make the identification:
\begin{equation}
T^{*}={\cal S}(T)^{t}~~~\mbox{and}~~~
(dT)^{*}={\cal S}(dT)^{t}=-(T^{-1})^{t}(dT)^{t}(T^{-1})^{t}~.
                                                         \label {i}
\end{equation}
One can check that the operation ${*}$ introduced in the last equation is the
involution of the Hopf
algebra $\delta _{{\cal H}}(GL_q(N))$. The ${*}$-Hopf algebra arising in
such a way is
nothing but the $\delta _{{\cal H}}$ extension $\delta _{{\cal H}}(U_q(N))$ of
the algebra $Fun(U_{q}(N))$.

Let us show that the quantum determinant $D$ for the
$GL_{q}(N)$
is not a central element of the extended algebra $\delta _{{\cal H}}(GL_q(N))$.
It is well known (\cite{Fad}) that $D$ can be written in the form:
\begin{equation}
D=\tr \mbox{\left(P^{-}(T\otimes T) \right)
=}\tr \mbox{\left( P^{-}T_{1}T_{2}P^{-} \right)}      ~,
                                                                   \label {d}
\end{equation}
where $P^{-}$ is a projector:
\begin{equation}
P^{-}=\frac{-PR+qI}{q+\frac{1}{q}}~,
                                                      \label {p}
\end{equation}
that can be treated as the quantum analog of symmetrizator in
$C^{n}\otimes C^{n}$. Then we have:
\begin{equation}
(dT)_{1}P^{-}_{23}T_{2}T_{3}P_{23}=P^{-}_{23}R_{12}^{-1}R_{13}^{-1}
T_{2}T_{3}(dT)_{1}R_{31}^{-1}R_{21}^{-1}P_{23}~.
                                                    \label {rop}
\end{equation}
For the projector $P^{-}$ we have the relation that follows from
definition (\ref{p}):
\begin{equation}
P^{-}_{23}R_{13}R_{12}=qP^{-}_{23}~.
                                                     \label {tor}
\end{equation}
Therefore (\ref{rop}) reduces to
$$\frac{1}{q}P^{-}_{23}T_{2}T_{3}(dT)_{1}R_{31}^{-1}R_{21}^{-1}P^{-}_{23}~.$$
The transposition of (\ref{tor}) gives
$$
R_{31}^{-1}R_{21}^{-1}P^{-}_{23}=\frac{1}{q}P^{-}_{23}~,
$$
where the fact was used that $(P^{-}_{23})^{t}=P^{-}_{23}$.
Now equation (\ref{rop}) takes the form:
$$(dT)_{1}P^{-}_{23}T_{2}T_{3}P^{-}_{23}=
\frac{1}{q^{2}}P^{-}_{23}T_{2}T_{3}P^{-}_{23}(dT)_{1}~.$$
Finally taking the trace we obtain:
\begin{equation}
dTD=\frac{1}{q^{2}}DdT~.
                                                           \label {det}
\end{equation}
Thus $D$ is not a central element of $\delta _{{\cal H}}(GL_q(N))$. Applying
the involution to
$D$ we find that $D$ is the unitary element of $Fun(U_{q}(N))$:
$$D^{*}D=DD^{*}=I.$$ These two facts play a crucial role
in constructing $*$-representations of $Fun(U_{q}(N))$ in a separable Hilbert
space.

\section{Type I representation of the Hopf algebra $~~~~~~$
$\delta _{{\cal H}}(Fun(U_{q}(2)))$} Let us consider the question about
${*}$-rep\-re\-sen\-ta\-tions of the Hopf algebra
$~~~\delta _{{\cal H}}(Fun(U_{q}(2)))$
in the separable Hilbert space ${\cal H}$.
Since $Fun(U_{q}(2))$ has the completion that is a $C^{*}$-algebra the
elements $T_{ij}$ can be realized as bounded operators in ${\cal H}$.
Then  $\partial T_{ij}$ are unbounded operators. It can be proved by using
(\ref{det}).  Let us suppose that $\partial T_{ij}$ is a bounded operator
and $q<1$.  Then the norm $|| \partial T_{ij} ||$ is defined and
$$D^{\dagger}\partial T_{ij}D=\frac{1}{q^{2}}\partial T_{ij}~.$$ This
allows one to write $$| | D^{\dagger}\partial T_{ij}D| | =| | \partial
T_{ij}D| | \leq | | \partial T_{ij} | | | | D | | =| | \partial T_{ij} | |
.$$ So we obtain $$\frac{1}{q^{2}}| | \partial T_{ij} | | \leq | |
\partial T_{ij}| | $$ and therefore $1/q^{2}\leq 1$ that contradicts to
$q<1$.

It is difficult to begin with constructing representations
of the algebra
(\ref{cl}), (\ref{5}), (\ref{6}) since the involution condition for
$\partial T_{ij}$ is rather complicated (we will come back to this question
in the next section).
But this difficulty can be solved
by introducing
the new set of generators $L=\partial TT^{-1}$ on which the action
of the involution is simple. From (\ref{cl}), (\ref{5}) one can
deduce the defining relation for $T$ and $L$:
\begin{equation}
R_{12}L_{1}R_{21}T_{2}=T_{2}L_{1}~~.
                                                    \label {pair}
\end{equation}
The pair $(T,L)$ is called the $(T,L)$-pair \cite{AV}.
The involution property for $L$ is $$L^{\dagger}=-L~.$$

Now we are going to construct the special representation of the $(T,L)$-pair
in a Hilbert space for the case $T\in Fun(U_{q}(2))$.
Let $T=| | t_{ij} | |$ be
the matrix of the form:
$$
T=\left(\begin{array}{cc}a & b \\ c & d\end{array}\right),
$$
the inverse of which is
$$
T^{-1}=D^{-1}
\left(\begin{array}{rr} d & -\frac{1}{q}b \\-qc & a\end{array}\right).
$$
Here $D=ad-qbc~$ is the quantum determinant. Taking into account the involution
one can write:
$a^{*}=D^{-1}d$ and $c^{*}=-\frac{1}{q}D^{-1}b$. $L$ is the matrix:
$$L=\left(\begin{array}{cc}\omega _{0} & \omega _{+} \\
\omega _{-} & \omega _{1}\end{array}\right)~,$$ where the entries obey the
following involution relations $\omega _{0}^{*}=-\omega _{0}$,
$\omega _{+}^{*}=-\omega _{-}$, $\omega _{1}^{*}=-\omega _{1}$, i.e.
$\omega _{0}$ and $\omega _{1}$ are antihermitian. The explicit form
of the permutation relations for the $(T,L)$-pair is
$$a\omega _{0}=q^{2}\omega _{0}a,~~~~c\omega _{0}=\omega _{0}c,$$
\begin{equation}
b\omega _{0}=q^{2}\omega _{0}b,~~~~d\omega _{0}=\omega _{0}d,
                                                    \label {p1}
\end{equation}

$$a\omega _{+}=
q\omega _{+}a,~~~~ c\omega _{+}=q\omega _{+}c+\mu \omega _{0}a,$$
\begin{equation}
b\omega _{+}=q\omega _{+}b,~~~~d\omega _{+}=q\omega _{+}d+\mu \omega _{0}b,
                                                    \label {p2}
\end{equation}

$$c\omega _{-}=
q\omega _{-}c,~~~~a\omega _{-}=q\omega _{-}a+\mu \omega _{0}c,$$
\begin{equation}
d\omega _{-}=q\omega _{-}d,~~~~b\omega _{-}=q\omega _{-}b+\mu \omega _{0}d,
                                                    \label {p3}
\end{equation}

$$a\omega _{1}=
\omega _{1}a+\mu c\omega _{+},~~~~b\omega _{1}=\omega _{1}b+\mu d\omega _{+},$$
\begin{equation}
c\omega _{1}=
q^{2}\omega _{1}c+
q\mu \omega _{-}a,~~~~d\omega _{1}=q^{2}\omega _{1}d+q\mu \omega _{-}b,
                                                    \label {p4}
\end{equation}
where $\mu=q-\frac{1}{q}$.

Let $\pi $ be a ${*}$-representation of the algebra $Fun(U_{q}(N))$
in the separable Hilbert space $l_2$. The operators $\pi (t_{ij})$ are supposed
to be continuous ones. In \cite{LS,Ko} it was proved that every irreducible
${*}$-representation $\pi$ of $Fun(U_{q}(N))$ is unitary equivalent
to the one of the following two series:
\begin{enumerate}
\item One dimensional representations  $\xi_{\psi}$ given by the formulae:
\begin{displaymath}
      \xi_{\psi}(a)=e^{i\psi}~~~~~,~~~~~\xi_{\psi}(c)=0~~,~~\psi\in
  R/{2\pi Z}~~.
\end{displaymath}
\item Infinite
dimensional representations $\rho_{\phi,\theta}$ in a Hilbert space
with orthonormal basis $\{ e_n\} _{n=0}^{\infty}$:
\begin{displaymath}
   \rho_{\phi,\theta}(a)e_0=
   0~~,~~\rho_{\phi,\theta}(a)e_n=e^{i(\theta+\phi)}\sqrt{1-e^{-2nh}}e_{n-1}
  ~~,~~\rho_{\phi,\theta}(c)e_n=e^{i\theta}e^{-nh}e_n
\end{displaymath}
\begin{equation}
\rho_{\phi,\theta}(d)e_n=\sqrt{1-e^{-2(n+1)h}}e_{n+1}
  ~~,~~\rho_{\phi,\theta}(b)e_n=-e^{i\phi}e^{-(n+1)h}e_n
                                                           \label{pred}
\end{equation}
Here $\theta, \phi \in [0,2\pi)$, $~ q=e^{-h}$.
\end{enumerate}

Thus for $Fun(U_{q}(N))$ the set $\hat{{\cal F}}$ of equivalence classes of
irreducible unitary representations consists of two separate
components each of these is numerated by continuous parameters
$\theta,\phi \in T^2=S^1\times S^1$ playing the role of coordinates. We shall
concentrate our
attention on the infinite-dimensional component since only a trivial
representation of differentials corresponds to
one-dimensional representations of $U_q(2)$.

The straightforward
algebraic consideration shows that it is impossible to realise the operators
$\omega _{0}, \omega _{+}, \omega _{-}, \omega _{1}$ in the space of
an irreducible representation of the algebra $Fun(U_{q}(2))$. This means that
one have to extend the space of representation or in other words to work with
reducible representations of $Fun(U_{q}(2))$.
It turns out that a suitable construction deals with a direct integral
of Hilbert spaces. Let us consider the Hilbert space ${\cal H}$ of
functions on a circle taking value in $l_2$. It is known  \cite{Dix} that
there exists the canonical isomorphism ${\cal H}=l_2\otimes {\cal
L}_2(S^1)$ and $$ {\cal H}=\int_{S^1}{\cal H}(\phi)d\phi, $$ where ${\cal
L}_2(S^1)$ is the space of square integrable functions on a circle obeying the
condition:  $$ \int ^{\pi}_{-\pi}| f(\phi)| ^2d\phi< \infty~, $$ for any
$f\in {\cal L}_2(S^1)$ and ${\cal H}(\phi)=l_2$. Now the reducible
representation of  $Fun(U_{q}(2))$ in $\cal {H}$ can be defined in the
following manner:  \begin{displaymath} \hat{a}(e_0\otimes f)=0,
\end{displaymath}
\begin{displaymath}
\hat{a}(e_n\otimes f)=\sqrt{1-e^{-2nh}}e_{n-1}\otimes e^{i(\theta+\phi)}f,
\end{displaymath}
\begin{equation}
\hat{d}(e_n\otimes f)=\sqrt{1-e^{-2(n+1)h}}e_{n+1}\otimes f,
                                    \label{hpr}
\end{equation}
\begin{displaymath}
\hat{b}(e_n\otimes f)=-e^{-(n+1)h}e_n \otimes e^{i\phi}f,
\end{displaymath}
\begin{displaymath}
\hat{c}(e_n\otimes f)=e^{-nh}e_n\otimes e^{i\theta}f,
\end{displaymath}
where
the operators $\hat{a},\hat{b},\hat{c},\hat{d},$
correspond to $a,b,c,d$.
On putting
$\phi$  equal to some value $\phi _0$ the irreducible representation
$\pi_{\theta,\phi_0}$ of $Fun(U_q(N))$ stands out.

The scalar product in ${\cal H}$ is given by
\begin{equation}
<(e_n\otimes f),(e_m\otimes g)>=(e_n,e_m)\int_{-\pi}^{\pi}f\bar{g}d\phi~.
                                                  \label {sk}
\end{equation}

Introduce the following hermitian
operator $K$  defined on a dense region in ${\cal L}_2(S^1)$:
\begin{equation}
\left(Kf\right)(\theta,\phi)=
\sum_{n}a_{nm}q^{2n}\gamma^{n}=
\left(e^{-2ih\frac{\partial}{\partial \phi}
} f\right)(\theta,\phi), \label{cr} \end{equation} where $f=
\sum_{n}a_{n}\gamma^{n}$ is an arbitrary element of
${\cal H}, \gamma=e^{i\phi}$.

Now let us take $\omega _{0}$ to be the operator in ${\cal H}$:
\begin{equation}
\omega_{0}(e_n\otimes f)=ie_{n}\otimes e^{-2ih\frac{d}{d\phi}}f.
                                                    \label {rip}
\end{equation}
Then $\omega _{0}$ is antihermitian as it is required.
The permutation relations of $\omega_{0}$ with the operators $\hat{a},\ldots ,
\hat{d}$ are precisely (\ref{p1}). For the operator $\omega_{+}$ one can choose
the realization:
\begin{equation}
\omega_{+}(e_n\otimes f)=-ie^{(n+2)h}\sqrt{1-e^{-2nh}}e_{n-1}
\otimes e^{i\phi}e^{-2ih\frac{d}{d\phi}}f.
                                                  \label {om}
\end{equation}
Introducing formally the inverse operators $\hat{b}^{-1}$ and $\hat{c}^{-1}$:
$$
\hat{b}^{-1}(e_{n}\otimes f)=-e^{(n+1)h}e_{n}\otimes e^{-i\phi}f,
$$
$$
\hat{c}^{-1}(e_{n}\otimes f)=e^{nh}e_{n}\otimes e^{-i\theta}f,
$$
it is possible to
rewrite $\omega_{+}$ in terms of $\hat{a},\ldots,\hat{c}^{-1}$:
\begin{equation}
\omega_{+}=-q^{-3}\widehat {c^{-1}}\hat{a}e^{-2ih\frac{d}{d\phi}}=
-q^{-3}\hat {c}^{-1}\hat{a}\omega_{0}.
                                                    \label {cor}
\end{equation}
By straightforward calculations one can check the fulfilment of the
permutation relations (\ref{p2}) on a dense region where all operators
coming in (\ref{cor}) are defined. Taking $\omega_{-}$ to be the hermitian
conjugation of $\omega_{+}$ with respect to the scalar product (\ref{sk}):
\begin{equation}
\omega_{-}=\hat{d}\hat{b}^{-1}\omega_{0},
                                                        \label {om1}
\end{equation}
we find that the relations (\ref{p3}) are satisfied.

Now the question arises how to find the operator $\omega_{1}$ that must be
antihermitian and obey (\ref{p4}). We choose for $\omega_{1}$ the
following ansatz:
\begin{equation}
\omega_{1}={\cal P}(\hat{a},\ldots,\hat{c}^{-1})\omega_{0},
                                                        \label {an1}
\end{equation}
where ${\cal P}(\hat{a},\ldots,\hat{c}^{-1})$ is a polynomial in
$\hat{a},\ldots,\hat{c}^{-1}$. Then the first line in (\ref{p4}) reads:
\begin{equation}
\hat{a}{\cal P}=\frac{1}{q^2}{\cal P}\hat{a}-
\frac{\mu}{q^3}\hat{a}~~,~~
\hat{b}{\cal P}=\frac{1}{q^2}{\cal P}\hat{b}-
\frac{\mu}{q^3}\hat{d}\hat{c}^{-1}\hat{a},
                                                                 \label {pol}
\end{equation}
and the second
one is achieved from the first by hermitian conjugation. Equations
(\ref{pol})
have the simple
solution ${\cal P}=-\frac{1}{q^2}\hat{b}^{-1}\hat{d}\hat{c}^{-1}\hat{a}$.
Therefore $\omega_{1}$ takes the form:
\begin{equation}
\omega_{1}=-\frac{1}{q^2}\hat{b}^{-1}\hat{d}\hat{c}^{-1}\hat{a}
                                                         \label{really}
\end{equation}
The found operators $\omega_{0},\omega_{+},\omega_{-},\omega_{1}$ combined
with (\ref{hpr}) give ${*}$-representation of the $(T,L)$-pair. Coming back
to the derivatives $\partial T=LT$ we see that
\begin{equation}
\begin{array}{ll} \partial a=
0, & \partial b=-\frac{1}{q^3}c^{-1}D\omega_{0}, \\
\partial c=0, & \partial d=-\frac{1}{q^3}d(bc)^{-1}D\omega_{0}.
\end{array}
                                                         \label{rell}
\end{equation}
This means that our differentials
can be treated as elements of the algebra $Fun(U_q(2)$
which is extended by adding the new element $\omega_0$ provided that the
elements $b$ and $c$ are invertible.  The equalities  $\partial a=0$ and
$\partial c=0$ seem rather restrictive and give a hint that
other ${*}$-representations for which $\partial a\neq 0$, $\partial
c\neq0 $ should also exist.  In the next section we will construct one of
such examples.

\section{Type II ${*}$-representation of
$\delta _{{\cal H}}(Fun(U_{q}(2)))$}
The permutation relations between the elements
$a,b,c,d$ and their derivatives follow from (\ref{5}). In particular we
have:  \begin{equation} \begin{array}{ll} a(\partial a)=q^2(\partial a)a,&
c(\partial a)=q(\partial a)c, \\ b(\partial a)=q(\partial a)b, &
d(\partial a)=(\partial a)d.  \end{array} \label{pem1} \end{equation} Let
us consider now the Hilbert space of
squire integrable functions on a torus $T^2=S^1\times S^1$ taking value in
$l_2$.  The scalar product in ${\cal H}$ has the form:  \begin{equation}
(f,g) =\int _{T^2} <f(\theta,\phi),g(\theta,\phi)> d\theta d\phi ,
\label{sc1} \end{equation} where $<,>$ is a scalar product in $l_2$. As in
the previous section the reducible representation of $Fun(U_q(2))$ in
${\cal H}$ can be defined by the formulas (\ref{hpr}) where $\theta $ is
no longer a parameter and $f=f(\theta , \phi)\in {\cal H}$.

Note that equations (\ref{pem1}) are compatible with the condition
$\partial a=(\partial a)^{*}$. This allows one to choose for $\partial a$ the
simple realization by the hermitian unbounded operator:
\begin{equation}
\left(\partial \hat{a}\right)(\theta,\phi)=
\sum_{nm}a_{nm}q^{n+m}\gamma_{1}^{n}\gamma_{2}^{m}=
\left(e^{-ih\frac{\partial}{\partial \phi}-ih\frac{\partial}{\partial \theta}}
f\right)(\theta,\phi),
\label{h1} \end{equation}
where $f=
\sum_{nm}a_{nm}\gamma_{1}^{n}\gamma_{2}^{m}$ is an arbitrary element of
${\cal H}$, $\gamma_1=e^{i\theta}$, $\gamma_2=e^{i\phi}$ and $\partial
\hat{a}$ is the operator that corresponds to $a$.

One can go further and require the condition
$\partial a=(\partial a)^{*}$ to be consistent with the involution
(\ref{i}) whose explicit form is \begin{equation} (\partial
a)^{*}=-q^2(D^{-1})\left(q^2d^2(\partial a)-bd(\partial c)-qdc(\partial
                    b)+\frac{1}{q}bc(\partial d)\right),
                                                               \label{i1}
\end{equation}
\begin{equation}
(\partial d)^{*}=
-q^2(D^{-1})\left((q^2ad-D)
(\partial a)-qab(\partial c)-qac(\partial b)+a^2(\partial d)\right),
                                                               \label{i2}
\end{equation}
\begin{equation}
(\partial b)^{*}=
-q^2(D^{-1})\left(-q^3cd
(\partial a)+qad(\partial c)+q^2c^2(\partial b)-ac(\partial d)\right),
                                                               \label{i3}
\end{equation}
\begin{equation}
(\partial c)^{*}=
-q^2(D^{-1})\left(-q^2db(\partial a)
+b^2(\partial c)+\frac{1}{q}ad(\partial b)-\frac{1}{q}ba(\partial d)\right).
                                                               \label{i4}
\end{equation}

I this way we can express $\partial d$ in terms of $\partial a$, $\partial b$
and $\partial c$:
\begin{equation}
\partial d=q(bc)^{-1}\left(-(q^2d^2
+\frac{1}{q^2}D^2)(\partial a)+bd(\partial c)+qdc(\partial b)\right).
                                                               \label{id}
\end{equation}
Therefore to specify $\partial d$ we need the explicit realization of $\partial
b$
and $\partial c$ that are compatible with (\ref{i3}), (\ref{i4}). The
permutation relations between $\partial b$, $\partial c$ and the elements of
$Fun(U_q(2))$ are
\begin{equation}
\begin{array}{ll}
a(\partial c)=q^2(\partial c)a+q\mu (\partial c), &
c(\partial c)=q^2(\partial c)c, \\
b(\partial c)=(\partial c)b+\mu (\partial a)d , &
d(\partial c)=(\partial c)d,
\end{array}
                                                                \label{pem2}
\end{equation}
\begin{equation}
\begin{array}{ll}
a(\partial b)=q(\partial b)a+\mu b(\partial a), &
c(\partial b)=(\partial b)c+\mu d(\partial a), \\
b(\partial b)=q^2(\partial b)b, &
d(\partial b)=q(\partial b)d.
\end{array}
                                                                \label{pem3}
\end{equation}
To solve  equations (\ref{pem2}) and (\ref{pem3}) we take as in the
previous section:
$$
\partial b={\cal I} \partial a~~~,~~~\partial c={\cal J} \partial a
$$
where ${\cal I}$ and ${\cal J}$ are polynomials in $\hat{a},\ldots,\hat{d},
\hat{b}^{-1},\hat{c}^{-1} $ that should be defined. By simple computations
we find:
\begin{equation}
\partial \hat{b}=q\hat{c}^{-1}\hat{d} \partial a~~~,~~~\partial
\hat{c}=\hat{d}\hat{b}^{-1} \partial \hat{a}.
                                                        \label{bc}
\end{equation}
Then (\ref{id}) reduces to the form:
\begin{equation}
\partial \hat{d}=(\hat{b}\hat{c})^{-1}(q^3\hat{d}^2
-\frac{1}{q}D^{2})\partial \hat{a}.
                                                        \label{ddd}
\end{equation}
Using the found representation                        of the derivatives
(\ref{bc}),(\ref{ddd}) one can show that the permutation
relations for $\partial d$:
\begin{equation}
\begin{array}{ll}
a(\partial d)=(\partial d)a+\mu (\partial b)c+\mu b(\partial c), &
c(\partial d)=q(\partial d)c+\mu d(\partial c), \\
b(\partial d)=q(\partial d)b+\mu q(\partial b)d, &
d(\partial d)=q^2(\partial d)d.
\end{array}
                                                        \label{ss}
\end{equation}
are satisfied.
The conjugation of the operators $\partial b,\partial c,\partial d$ with
respect to the scalar product (\ref{sc1}) gives
\begin{equation}
\begin{array}{l}
(\partial b)^{*}=-qab^{-1}(\partial a), \\
(\partial c)^{*}=-\frac{1}{q^2}c^{-1}(a\partial a), \\
(\partial d)^{*}=q(a^2-1)(bc)^{-1}(\partial a).
\end{array}
                                                        \label{sss}
\end{equation}
The consistency of equations (\ref{sss}) with (\ref{i1})-(\ref{i4}) can be
easily checked. Hence the formulae (\ref{h1}),(\ref{bc}),(\ref{ddd}) give
the other ${*}$-representation of the $\delta _{{\cal H}}$-extension of
$Fun(U_q(2))$.
We refer to this representation as the type II.

\section{Two-dimensional Local Gauge Transformations}
Now having at hand an explicit representations of the $\delta _{{\cal H}}$-
extension of $U_q(N)$ we are able to construct
two-dimen\-sio\-nal infinitezimal local gauge transformations.
As was mentioned in Introduction it seems reasonable to identify
the representations of types I and II
 with derivations in two linear independent
directions of space-time. To construct an explicit realization of
two-dimensional differentials of $U_q(2)$ elements it is
convenient to introduce the operators $K_1$ and $K_2$ acting on $\cal{H}$
in the following manner:  \begin{equation} (K_1f)(\theta,\varphi)=
\sum_{nm}a_{nm}q^{2n}\gamma_{1}^{n}\gamma_{2}^{m}=
\left(e^{-2ih\frac{\partial}
{\partial\varphi}}f\right)(\theta,\varphi),
                                           \label{ki}
\end{equation}
\begin{equation}
(K_{2}f)(\theta,\varphi)= \sum_{nm}a_{nm}q^{n+m}\gamma_{1}^{n}\gamma_{2}^{m}=
\left(e^{-ih\frac{\partial}{\partial\varphi}
-ih\frac{\partial}{\partial\theta}}f\right)(\theta,\varphi)
                                           \label{kii}
\end{equation}
where $\gamma_1=e^{i\theta}$, $\gamma_2=e^{i\phi}$ and $f\in {\cal H}$.
Then according to our conjecture we may regard the formulae (\ref{rell}) as
the derivatives of the elements of $U_q(2)$ with respect to the coordinate
$x^1$ on $R^2$, and the formulae (\ref{h1}),(\ref{bc}),(\ref{ddd})
as the derivatives in the
$x^2$ direction. Thus the decomposition of differentials on $U_q(2)$ over
the basis $\{ dx^i\}$ of differential forms on $R^2$ is
\begin{displaymath}
\delta a=K_{2}\otimes dx^2,
\end{displaymath}
\begin{equation}
\delta b=-\frac{1}{q^3}c^{-1}DK_{1}\otimes dx^1+qc^{-1}dK_{2}\otimes dx^2,
										  \label{dif}
\end{equation}
\begin{displaymath}
\delta c=db^{-1}K_{2}\otimes dx^2,
\end{displaymath}
\begin{displaymath}
\delta d=-\frac{1}{q^3}d(bc)^{-1}DK_{1}\otimes dx^1+(bc)^{-1}
\left(q^3d^2-\frac{1}{q}D^2\right)K_{2}\otimes  dx^2.
\end{displaymath}
(We use $\delta $ for exterior derivative rather than $d$ to
avoid misunderstanding with the element $d$).

Let us make a comment about equation (\ref{6}). Formally this equation
follows from (\ref{5}) and the identity $d^2=0$. However, we have not define
in the above construction an action of the operator $d$ on $dT$. Nevertheless,
$dT=\left(\begin{array}{ll}\delta a & \delta b \\
\delta c & \delta d\end{array}
\right)$
does satisfy the relation (\ref{6}). This is non-trivial since the right hand
side of (\ref{6})
contains two differentials $dx^1$ and $dx^2$ and is the result of
direct calculations.

The meaning of the
formulae (\ref{dif}) is that locally we have a two-parameter map
$R^2\rightarrow U_q(2)$ compatible with the bicovariant
differential calculus on $U_q(2)$. For example, one can write $$
b(x^1,x^2)=
b-\frac{1}{q^3}c^{-1}DK_{1}\otimes x^1+qc^{-1}dK_{2}\otimes x^2,
$$
then the usual derivations of $b$ with respect to $x^1$ or $x^2$ give
the operators on $\cal{H}$ with desirable properties
(\ref{5}),(\ref{6}). It  would be interesting to construct a global map
$R^2\rightarrow U_q(2)$ and this is the subject of further investigations.
$$~$$
{\bf ACKNOWLEDGMENT}
$$~$$
The authors are grateful to P.B.Medvedev and I.Volovich for interesting
discussions.
$$~$$

\end{document}